\documentclass[twocolumn,aps,floatfix,superscriptaddress]{revtex4-1}
\usepackage{lineno}
\usepackage[usenames,dvipsnames]{xcolor} 
\usepackage{graphicx}
\usepackage{textcomp}
\usepackage{float}
\usepackage{amsfonts}
\usepackage{amsmath}
\usepackage{amssymb}
\usepackage{enumerate}
\usepackage[normalem]{ulem}
\usepackage{color}
\usepackage{color,soul} 
\usepackage{multirow} 
\usepackage{rotating} 
\usepackage{verbatim}
\usepackage[sort&compress]{natbib}
\usepackage[english]{babel}
\usepackage{xr}
\begin{document}

\title{Spatial, spectral, temporal and polarisation resolved state tomography of light}

\author{Martin Pl\"{o}schner}
\thanks{Corresponding author; E-mail: m.ploschner@uq.edu.au}
\affiliation{School of Information Technology and Electrical Engineering, The University of Queensland, Brisbane, QLD 4072, Australia}

\author{Marcos Maestre Morote}
\affiliation{School of Information Technology and Electrical Engineering, The University of Queensland, Brisbane, QLD 4072, Australia}

\author{Daniel Dahl}
\affiliation{School of Information Technology and Electrical Engineering, The University of Queensland, Brisbane, QLD 4072, Australia}

\author{Mickael Mounaix}
\affiliation{School of Information Technology and Electrical Engineering, The University of Queensland, Brisbane, QLD 4072, Australia}

\author{Greta Light}
\affiliation{II-VI Incorporated, 48800 Milmont Dr., Fremont CA 94538, USA}

\author{Aleksandar Rakic}
\affiliation{School of Information Technology and Electrical Engineering, The University of Queensland, Brisbane, QLD 4072, Australia}

\author{Joel Carpenter}
\affiliation{School of Information Technology and Electrical Engineering, The University of Queensland, Brisbane, QLD 4072, Australia}

\begin{abstract}

\textbf{The ability to measure polarisation, spectrum, temporal dynamics, and spatial amplitude and phase of optical beams is essential to study fundamental phenomena in laser dynamics, telecommunications and nonlinear optics. Current characterisation techniques only apply in limited contexts. Non-interferometric methods typically lack access to spatial phase, while phase-sensitive approaches necessitate either an auxiliary reference source or an adequate self-reference, neither of which is universally available. Regardless of the reference, deciphering complex wavefronts of multiple co-propagating incoherent fields remains particularly challenging. Here, we harness the principles of quantum state tomography to circumvent these limitations. A full description of an unknown beam is retrieved by measuring its temporally and spectrally resolved density matrices for both polarisations, using a spatial light modulator to display projective holograms and a single-mode fibre to guide the collected signal to a high-speed photodiode and a spectrometer. Despite no spatial resolution of the detector and the intensity-only character of the collected signal, the method resolves multiple arbitrary spatial fields within a single beam, including their phase and amplitude, as well as their spatial coherence. Leveraging the coherence information unlocks unambiguous determination of the spectral and temporal evolution of mutually incoherent fields, even when these spectrally overlap or have an identical time delay. We demonstrate these hallmark features by characterising the rich spatiotemporal and spectral output of a vertical-cavity surface-emitting laser diode that has so far resisted full analysis using existing techniques.}

\end{abstract}

\maketitle

\section*{INTRODUCTION}
Information can be encoded in optical beams by design, either in space through sculpted complex wavefronts\cite{mosk_controlling_2012, dholakia_shaping_2011, rubinsztein-dunlop_roadmap_2017} or, in time, through tailored pulse envelopes\cite{weiner_femtosecond_2000}, with further multiplexing of information channels possible using the spectral and polarisation dimensions. Encoded information can also be a consequence of natural phenomena such as the nonlinear interaction of light with the gain medium of the laser cavity\cite{brejnak_boosting_2021, gensty_wave_2005, bittner_suppressing_2018}.  In both scenarios, optical beams can be generally composed of multiple incoherent spatial fields, with variable time delays and with numerous spectral peaks. It is particularly challenging to fully characterise the spatial phase of the beam in this most general case -- that is, where the beam is composed of multiple spatial components that may or may not be mutually coherent, may or may not exist at the same time, and may or may not have the same wavelength. This diversity of components carried by the beam makes decoding the complete information extremely difficult in the expanding frontier of applications ranging from imaging\cite{booth_adaptive_2014} to linear and nonlinear spatio-temporal beam-shaping\cite{mounaix_time_2020, wright_controllable_2015,wright_spatiotemporal_2017}.

Conventional optical beam analysis techniques with the ability to recover not only the spatial intensity\cite{nicholson_spatially_2008, gao_single-shot_2014} but also spatial phase and coherence either utilise an external reference to recover the optical signal characteristics\cite{yamaguchi_phase-shifting_1997, goodman_digital_1967, fontaine_laguerre-gaussian_2019} or work on a self-referencing principle\cite{kaiser_complete_2009, forbes_creation_2016}. In the external reference case, the spatial amplitude and phase of light is measured by interfering the unknown optical beam with an external local oscillator, which imposes limitations on the spectral band and bandwidth that can be probed, with a suitable high-quality local oscillator not always available in the spectral region of interest. The self-referencing approaches, such as modal analysis \cite{pinnell_modal_2020, flamm_mode_2012}, alleviate the need for a local oscillator by using part of the beam itself as a reference. In this case, the spatial components of the unknown beam are determined via a series of projective intensity measurements using spatial correlation filters. The filters interfere probed modal components with a pre-selected internal reference mode to recover complex superposition coefficients of all spatial modes that constitute the original beam. The modal analysis method allows profiling of single-wavelength wavefronts\cite{schulze_wavefront_2012, paurisse_complete_2012} and dynamic tracking of temporal mode instabilities in lasers at a camera-speed\cite{stutzki_high-speed_2011, duparre_-line_2005, schmidt_real-time_2011}; however, only if at least one mode is known \textit{a priori} to exist in the beam \cite{kaiser_complete_2009}. Often, there is no such prior knowledge, or indeed the beam has no single mode that can serve as an appropriate reference. For example, even for omnipresent optical sources, such as laser diodes, the spatial components of the beam typically occupy spectrally discrete and distinct locations, which means there is no spatial component that can be singled out as a suitable reference over the whole spectral range. Similarly, an optical beam modulated in time has a dynamic wavefront composed of various spatial modes, with none guaranteed to exist over the whole temporal range and therefore none suitable as a reference. This severely narrows the range of applicability of self-referencing techniques and renders them incapable to analyse arbitrary optical beams.

A conceptually similar approach to modal analysis that can measure a completely arbitrary state, either mixed or pure, exists in quantum optics. In the context of spatial optical fields, a mixed state represents multiple mutually spatially incoherent fields whereas a pure state describes a single coherent field. By obtaining a sequence of projective tomographic measurements that are sensitive to certain physical aspects of the state, such as spatial features or polarisation, a complete representation of the state can be recovered in the form of the density matrix\cite{lvovsky_continuous-variable_2009, toninelli_concepts_2019}. The density matrix  formalism is applied to study a myriad of quantum states, ranging from vibrational modes in molecules \cite{dunn_experimental_1995} to an analysis of quantum gates \cite{obrien_demonstration_2003}. A number of studies also used the technique to study the classical degrees of freedom of light individually, including in the spatial \cite{mclaren_entangled_2012, yang_using_2016, milione_determining_2015, milione_higher-order_2011, ji_high-dimensional_2019,dennis_swings_2017, agnew_tomography_2011}, polarisation \cite{r_sheppard_three-dimensional_2016,salvail_full_2013} and the time-frequency domain \cite{gil-lopez_universal_2021}. However, none of the existing methods provides access to all the degrees of freedom of light simultaneously and therefore cannot reveal the rich, complex dynamics of arbitrary optical fields.

In this work, we use a high-dimensional ($N = 21$) analogue of Stokes polarimetry\cite{milione_determining_2015, salvail_full_2013} (Fig.~\ref{Figure1}(a,b)) to reconstruct an unknown optical beam in space, time, spectrum and polarisation via a series of projective measurements. The projective measurements are performed using a tandem of a spatial light modulator (SLM) and a single-pixel detector in the form of single-mode fibre (SMF) as illustrated in Fig.~\ref{Figure1}(b). Each projection measurement corresponds to a pairwise interference between two potential spatial components (modes) of the unknown beam, with the full measurement interfering all the possible spatial components that exist within the beam. 

In stark contrast to the modal analysis approach that necessitates prior knowledge of a reference mode, our approach does not require a known reference state, and handles even scenarios with multiple references or no reference at all. Any spatial state that exists in any wavelength or temporal bin can be leveraged by Stokes method as a reference. If there is a mixture of incoherent states in a given bin, all the states are acting as independent references for all the other states. As a result, the Stokes method can spatially analyse the beam along the spectral and temporal axis, as long as there is detectable light in a given wavelength or temporal bin, and the projective measurements span the existing spatial states in that bin. 

The collection of light with the single pixel detector in the form of SMF is modular in terms of routing and splitting the light to specialised detectors that are optimised for temporal and spectral resolution, such as oscilloscopes and spectrometers. This provides significant advantages over camera-based approaches that have limited refresh rates to resolve ultra-fast temporal dynamics, and low pixel counts to provide sufficient number of spectral bins. Probing the whole transversal profile of the beam at once by a single-pixel detector also improves the detection efficiency and decreases noise compared to raster-scanning or camera-based approaches that only examine a limited subspace of the beam at any time \cite{edgar_principles_2019}. Interestingly, despite no spatial resolution of the detector, the spatial resolution of the measured field can be arbitrarily high, since the basis used for projections is numerically calculated. 


We illustrate the merit of the technique by analysing the beam of a modulated vertical-cavity surface-emitting laser (VCSEL) diode which supports multiple mutually incoherent spatial modes within the beam, and can be readily modulated in time. The rich spatiotemporal and spectral output of the VCSEL resists full analysis using existing techniques despite its popularity in diverse products such as LiDAR for autonomous vehicles, face recognition and datacoms. Moreover, VCSEL attracts considerable attention in high-power applications, mainly from the perspective of fundamental studies such as spatio-temporal instabilities on a sub-nanosecond timescale\cite{bittner_suppressing_2018} and wave-chaotic behaviour due to symmetry breaking in the laser cavity\cite{brejnak_boosting_2021}.

\section*{RESULTS}
\subsection*{Principle of projective measurements}
\begin{figure*}[ht!]
	\includegraphics [width=16.0cm] {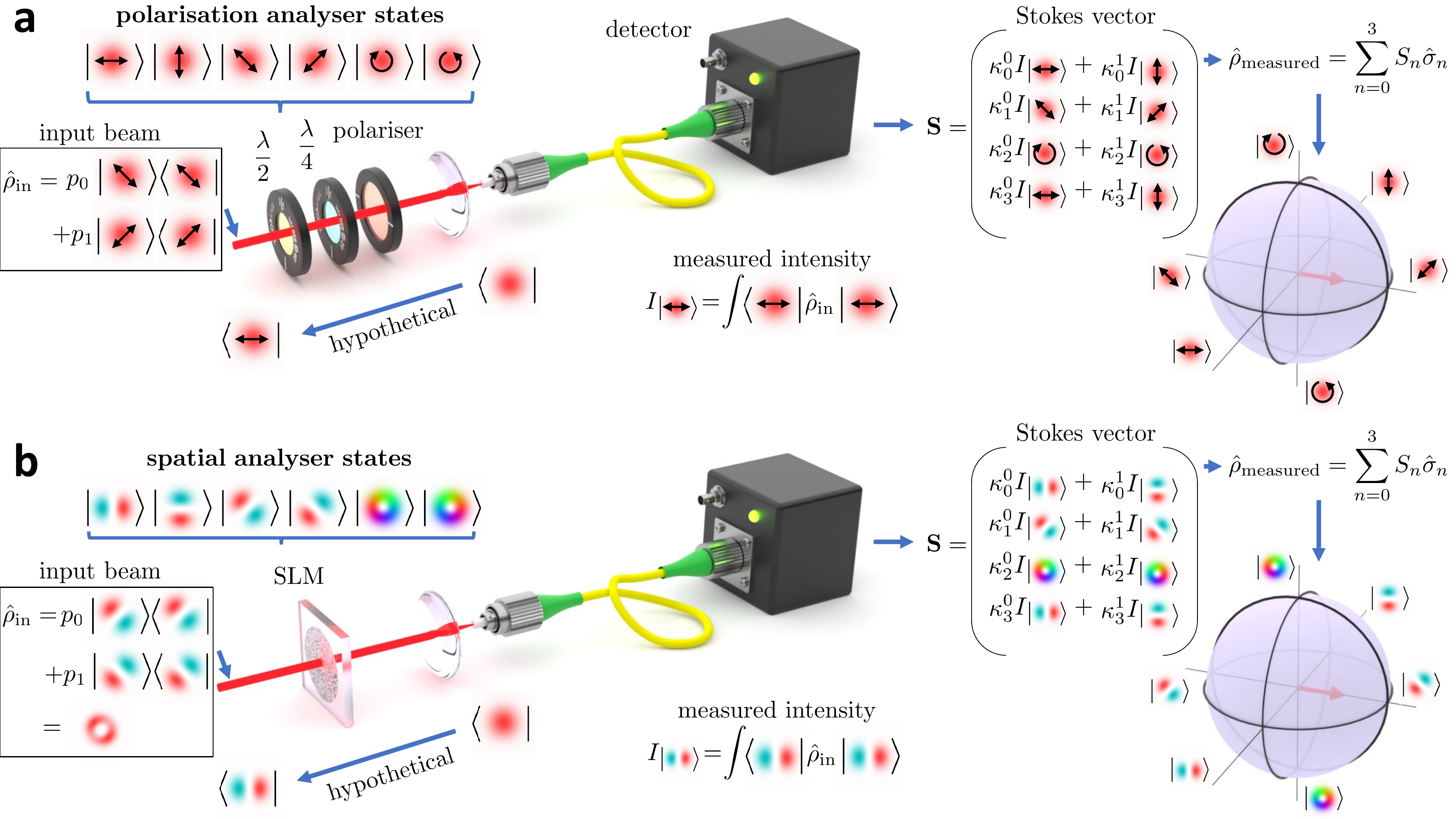}
	\caption{\textbf{Working principle of Stokes polarimetry and its spatial analogue.} An unknown pure or mixed state, described by density matrix $\hat{\rho}_{\text{in}}$, passes through a sequence of polarisition (\textbf{a}) or spatial (\textbf{b}) analyser states encoded on a spatial light modulator (SLM). Analyser states are determined by eigenvectors of Pauli matrices $\hat{\sigma}_n$ that represent all observables of a two dimensional Hilbert space. The intensity (expectation value) of a given observable is registered by a detector and a Stokes vector ($\mathbf{S}$) is reconstructed by weighting the measured intensities with eigenvalues $\kappa_n^m$ associated with eigenvectors of $\hat{\sigma}_n$. The sum of Pauli matrices weighted by the Stokes vector elements ($S_n$) determines the measured density matrix.  Measured mixed state can be graphically depicted within the volume of the Bloch sphere, whereas pure states reside on its surface\cite{milione_higher-order_2011}.}
	\label{Figure1}
\end{figure*}

In standard Stokes polarimetry illustrated in Fig.~\ref{Figure1}(a),
the unknown polarisation state of light, mathematically described by a density matrix $\hat{\rho}_{\text{in}}$, is interrogated by a sequence of polarisation analyser states that are defined by the orientation of the polarisation optics. In the depicted example, the power coupled into the SMF is equal to the expectation value of the unknown input state ($\hat{\rho}_{\text{in}}$) in the horizontal polarisation state. This horizontal polarisation state would exist at the input plane if the light was hypothetically launched from the SMF. Detecting the power with an SMF for multiple analyser states enables the recovery of the Stokes vector ($\mathbf{S}$), with its elements $S_n$ signifying weights of Pauli matrices $\hat\sigma_n$ in a sum that reconstructs the density matrix ($\hat{\rho}_{\text{measured}}$) of the unknown state. The density matrix encapsulates complete information about the unknown input state, including whether it represents a mixture of up to $N = 2$ states or a pure state. The situation is mathematically equivalent for spatial states (Fig.~\ref{Figure1}(b)), however, to perform the projective measurements experimentally in this case, we have to substitute the polarisation optics for a spatial light modulator, which acts as a reconfigurable spatial analyser state device. The power coupled into the SMF in this scenario is equal to the expectation value ($|\text{field overlap}|^2$) of the unknown input ($\hat{\rho}_{\text{in}}$) in the state defined by LP11 mode with lobes oriented horizontally, as illustrated in Fig.~\ref{Figure1}(b). We can again imagine that we hypothetically launch the light from the SMF towards the SLM which transforms the fundamental LP01 mode of the SMF into LP11 projective state at the source plane. We can mathematically overlap the source and the hypothetical LP11 field at the source plane but since no detector exists there the overlap value cannot be determined experimentally at this plane. However, the reciprocity of light ensures conservation of field overlap value in the system, which in turn allows its experimental recovery at the SMF plane (more details in Supplementary Note 4).

\subsection*{Experimental setup}
\begin{figure*}[ht!]
	\includegraphics [width=16.0cm] {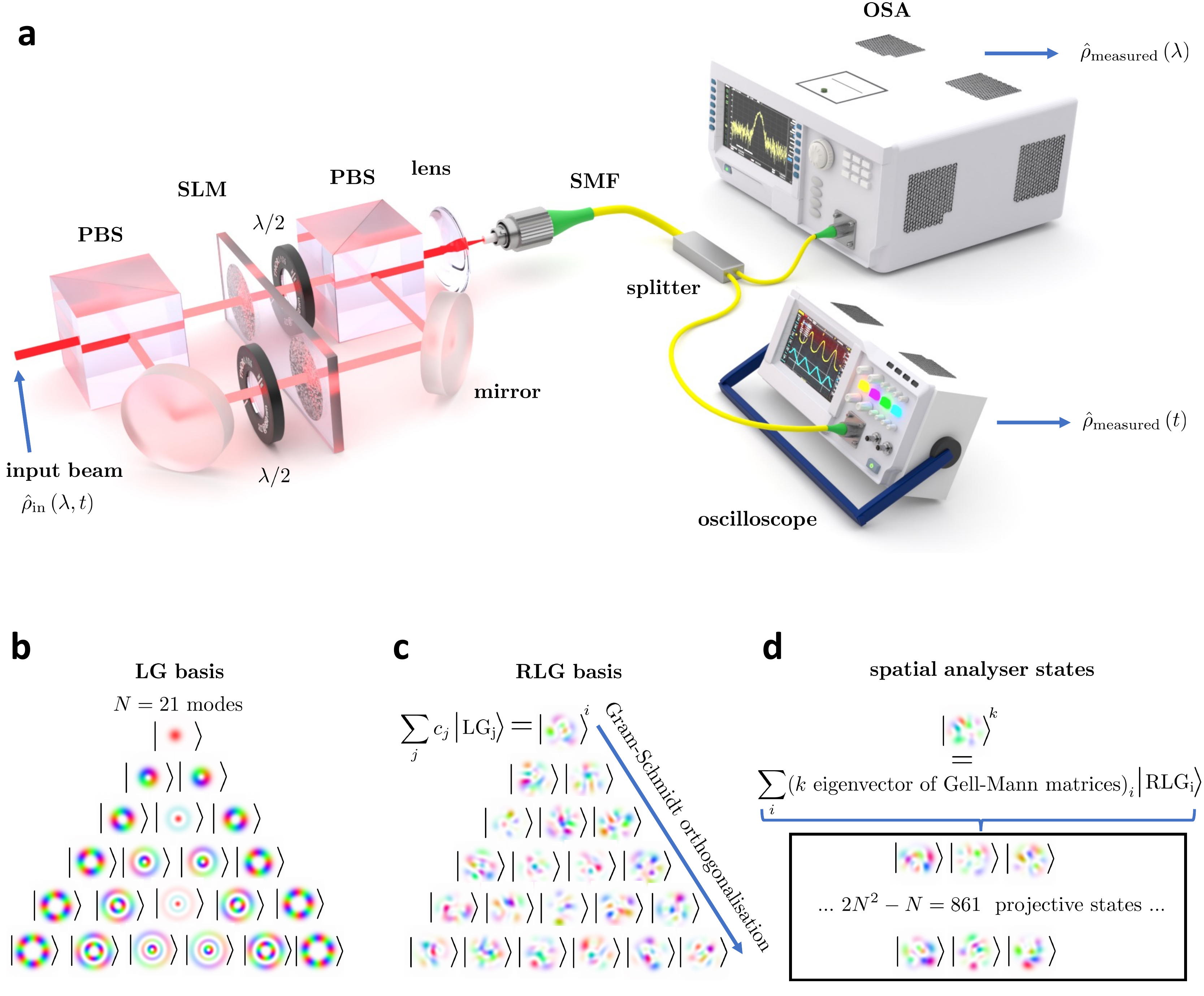}
	\caption{\textbf{Spatial, spectral and temporal state tomography} (\textbf{a}) The unknown state of light described by density matrix $\hat{\rho}_{\text{in}}\left(\lambda,t\right)$ is split into two optical paths by a PBS to allow independent interrogation of horizontal and vertical polarisation of the input state by the SLM. The recombined light is coupled into an SMF and guided into an oscilloscope and OSA for the analysis of the temporal and spectral degree of freedom. PBS - polarisation beam splitter, SLM - spatial light modulator, $\lambda/2$ - half-waveplate, SMF - single-mode fibre, OSA - optical spectrum analyser. A complex superposition of Laguerre-Gauss (LG) spatial states (\textbf{b}), given by random complex coefficients $c_j$, is used to construct one random Laguerre-Gauss (RLG) spatial state (\textbf{c}). The Gram-Schmidt orthogonalisation process generates the remaining states in the RLG basis (\textbf{c}). The k-th spatial analyser state (\textbf{d}) is constructed as a complex superposition of RLG spatial states with the complex superposition coefficients given by the k-th unique eigenvector of the Gell-Mann matrix set.}
	\label{Figure2}
\end{figure*}

Fig.~\ref{Figure2}(a) shows a simplified schematic of the experimental setup.
The unknown, pure or mixed state of light that generally depends on wavelength and time ($\hat{\rho}_{\text{in}}\left(\lambda, t\right)$) is separated by a polarisation beam splitter (PBS) into two paths to allow independent interrogation of vertical (V) and horizontal (H) polarisation by a single, large-area SLM that handles both polarisations. The orthogonal polarisation paths are recombined with another PBS and coupled into an SMF. The SMF subsequently routes the light to an optical spectrum analyser (OSA) and oscilloscope to analyse the input state with respect to wavelength and temporal degrees of freedom. More detailed experimental setup configuration can be found in Supplementary Note 1, along with the calibration and alignment procedures of the system in Supplementary Note 2.

The unknown state, represented by the density matrix ($\hat{\rho}_{\text{in}}\left(\lambda, t\right)$), can have a spatial profile with fine features due to the existence of multiple coherent and incoherent spatial fields at each wavelength and time. In order to extend the technique to more complicated spatial fields with more features, necessitates the extension of Stokes formalism from the limited two-dimensional case (Fig.~\ref{Figure1}(b)), where the analyser states are only sensitive to two modal components, into much higher dimension, until the Stokes states span all the modal components present in the beam. In our experimental case, a Stokes space dimension of $N = 21$ satisfies this condition, and actually exceeds it. When no \textit{a priori} information about the beam is known, the Stokes dimensionality $N$ should be chosen with some redundancy, at the price of carrying more projective measurements.

The natural generalisation of Pauli matrices, used for the reconstruction of the density matrix in two-dimensional scenario, into higher dimensions is facilitated by the generalised Gell-Mann matrices\cite{gell-mann_symmetries_1962} that span the space of observables of an $N$-dimensional Hilbert space. A vector in such an $N$-dimensional Hilbert space, specifically its vector elements, can represent complex superposition coefficients of an optical field in any orthogonal spatial basis of interest. The choice of this basis can be arbitrary. To demonstrate that the tomographic measurement can be performed using a random orthogonal spatial basis, we generate the spatial analyser states in the following fashion. We take as a starting point a Laguerre-Gaussian (LG) basis (Fig.~\ref{Figure2}(b)), with the waist of $w_0 = 3\, \rm{\mu m}$ reflecting the approximate extent of the investigated beam, and generate a single new spatial state composed of a random complex superposition of $N$ states in the LG basis (Fig.~\ref{Figure2}(c)). We subsequently apply the Gram-Schmidt orthogonalisation process to create the remaining $N-1$ orthogonal random spatial modes (Fig.~\ref{Figure2}(c)) to form the full random Laguerre-Gauss (RLG) basis with the dimension $N$. 

The number of modes $N$ in the initial selected basis should reflect the expected level of spatial details in the probed system. For instance, the spatial analyser states can sense high spatial frequency components only if the initial Stokes basis and the related random basis contains such high spatial frequency components. The choice of LG basis is convenient in cases with expected rotational symmetry, but the method is not limited to LG basis and works with Hermite-Gauss, position, Bessel, Hadamard or Fourier basis as per application needs. 

To generate the spatial analyser states that perform the tomographic measurements, we first find a set of unique eigenvectors of the Gell-Mann matrices. The vector elements of each eigenvector represent the complex superposition coefficients in the RLG basis and make up a given spatial analyser state (Fig.~\ref{Figure2}(d)). By sequentially applying the spatial analyser states using the SLM and measuring the intensity as a function of wavelength and time, using the OSA and oscilloscope respectively, we can reconstruct the generalised Stokes vector as both a function of wavelength and time. A blueprint for the calculation of $2N^2 -N = 861$ analyser masks for Stokes dimension of $N=21$ that generate the projective states, along with the mask alignment requirements is in Supplementary Note 2. Similar to the formalism presented in Fig.~\ref{Figure1}(b), by summing the Gell-Mann matrices with weights given by the Stokes vector, we reconstruct the spectrally and temporally resolved density matrices of the unknown beam. The eigenvectors and eigenvalues of the reconstructed density matrices (expressed in the RLG basis), are the spatial eigenstates and their probabilities in the investigated wavelength or temporal bin. Since both the spectrum and the time trace are collected in a single sweep for one specific spatial analyser state, only $2N^2 -N$ measurements in total are required for the full reconstruction of the spatial, spectral and temporal information of the beam. A detailed theory of the high-dimensional Stokes analysis is presented in Supplementary Note 3.

\subsection*{Spatio-spectral analysis}
\begin{figure*}[ht!]
	\includegraphics [width=16cm] {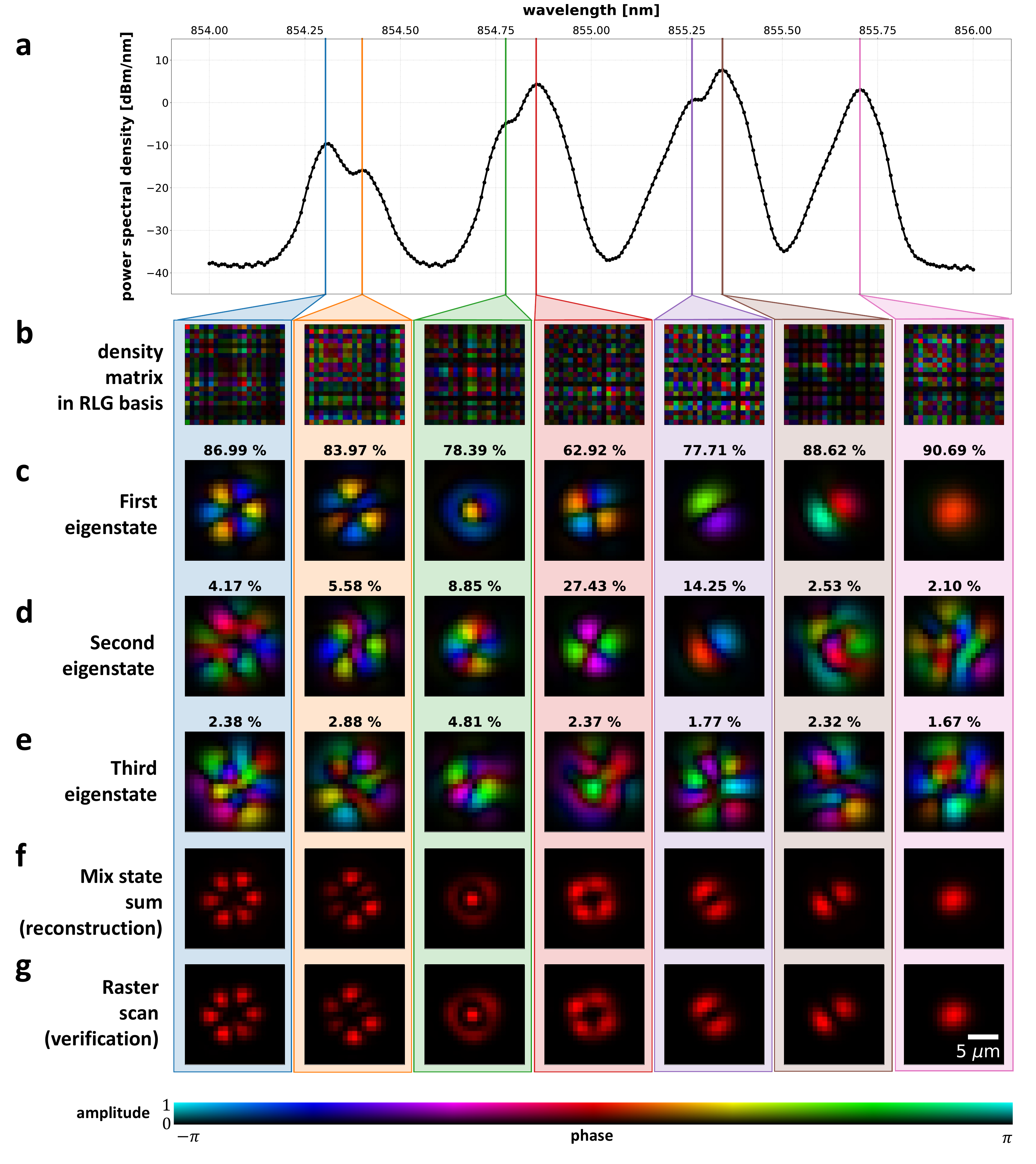}
	\caption{\textbf{Spatio-spectral analysis of VCSEL (H polarisation)} (\textbf{a}) The spectrum has multiple spectral peaks, each corresponding to one or more spatial modes. (\textbf{b}) The measured density matrix expressed in RLG basis for colour-coded spectral peaks. (\textbf{c-e}) The first three most dominant spatial eigenstates and their corresponding probabilities as reconstructed from the density matrix (\textbf{b}) using the RLG basis. (\textbf{f}) The probability-weighted intensity sum of all reconstructed eigenstates obtained via state tomography perfectly matches the intensity profiles (\textbf{g}) obtained by raster scanning the VCSEL beam over SMF by adding tilt on the SLM and collecting the spectrum for each tilt value. The measured density matrix along with the reconstructed spatial eigenstates for each wavelength is in Supplementary Media 1. The results for the V polarisation are presented in Supplementary Note 5 and Supplementary Media 2.}
	\label{Figure3}
\end{figure*}

We chose the vertical-cavity surface-emitting laser (VCSEL) diode as the unknown state $\hat\rho_{\text{in}}\left(\lambda, t\right)$ to demonstrate the method. VCSEL is a spatially, spectrally and temporally interesting source that represents a class of optical beams difficult to analyse using existing techniques. VCSEL cavity supports several spatial modes, with distributed spectral peaks (Fig.~\ref{Figure3}(a)).
The position of the spectral peaks and the number of spatial modes varies based on the VCSEL driving conditions like the bias current and the temperature (see Supplementary Note 7 for modally resolved light-current curves and details of driving conditions). The number of spatial modes and their relative power impacts the overall spatial profile of the VCSEL, which influences its performance in applications such as face recognition and 3D scene scanning for augmented reality \cite{liu_vertical-cavity_2019}. The ability to measure the spatio-spectral profile can provide useful feedback during VCSEL engineering and fabrication to improve the VCSEL performance in those applications. 

In this section, we use the measured spectrally resolved density matrix to distinguish the amplitude and phase of each spatial mode for each spectral slice. Uniquely, the density matrix unlocks the ability to resolve multiple mutually incoherent modes within the same spectral bin, which typically occurs in situations when spectral peaks of modes are closely packed and thus beyond the resolution of the spectrometer. We confirm the validity of the results by comparing the probability weighted sum of reconstructed mode intensities with the intensity profiles obtained from a simple raster-scan, with results obtained very similar in both cases. 

To perform the spatio-spectral analysis, we first collect the spectrum for each spatial analyser state in Fig.~\ref{Figure2}(d) and find the Stokes vector and the density matrix for each wavelength bin (resolution bandwidth of $70\,{\rm pm}$ given by the spectrometer) in the RLG basis. The measured density matrix at selected spectral peaks is depicted in Fig.~\ref{Figure3}(b). The eigenvalues and eigenvectors of the density matrix represent the probabilities and the spatial states at each wavelength, respectively. We note that the acquired density matrix should theoretically be positive-semidefinite to ensure positive probability for each supported state. However, due to measurement induced noise, this is not generally the case, and we have to apply Higham algorithm that finds the nearest positive semidefinite matrix in the Frobenius norm sense \cite{higham_computing_1988}. 

The mutually incoherent spatial states with the three highest probabilities at each spectral peak of Fig.~\ref{Figure3}(a) are shown in Fig.~\ref{Figure3}(c-e). The spatial states were constructed from the eigenvectors of the density matrix, with eigenvectors representing the complex superposition coefficients in the RLG basis. Despite the randomness of the RLG spatial basis, the reconstructed modes have the well-known linearly polarised (LP) mode spatial profiles typical for a circular laser cavity. Some spectral peaks have a significant second dominant state (Fig.~\ref{Figure3}(d)) and another weaker third state (Fig.~\ref{Figure3}(e)). This is most pronounced for the spatial modes corresponding to LP02 (green column), LP21 (red column) and LP11 (purple column) but also faintly visible for LP31 (orange column). In the LP11 case, the second dominant state is present simply due to leakage of the powerful neighbouring LP11 mode with a perpendicular orientation of lobes (brown). In the LP21 case, the wavelength bin contains a mixture of mutually incoherent states that are spectrally very close, within the $70\,{\rm pm}$ resolution of our spectrometer. Remarkably, even though we cannot spectrally resolve the modes, we can utilise the density matrix to recover not only the individual spatial profiles in the mixture but also their relative power. The ability of the technique to resolve spectrally overlapping spatial modes is one of its hallmark features. On the basis of mutual incoherence, it enables spatial differentiation of features beyond the spectral resolution. We explore this compelling attribute of Stokes analysis in detail in Supplementary Note 8, where we intentionally decrease the resolution bandwidth of the OSA to $10\,{\rm nm}$, fully spectrally mixing the spatial modes in a single detected wavelength bin, yet the Stokes analysis yields the correct spatial states and their relative powers in the mixture. For the LP21 case, the recovered probabilities in the mixture are $63\%$ for the LP21a and $27\%$ for the LP21b state. Apart from the two dominant modes totalling $90\%$ of power in the investigated wavelength bin, there is also approximately $2\%$ of LP02 mode as visible on a partially distorted spatial profile in Fig.~\ref{Figure3}(e). The $8\%$ of unaccounted power distributes among the remaining 18 spatial eigenstates (not displayed) of the density matrix, all of them with a random, speckle-like spatial profile. There is no physical reason to expect that such random spatial states exist in the laser cavity. We instead associate the $8\%$ of the remaining power to noise that inevitably occurs during the hundreds of sequential projections, either due to power fluctuation of the VCSEL or the thermally induced spectral offset.

The noise not only affects the amount of light that cannot be attributed to physical spatial modes but also influences the reconstruction fidelity of the spatial profile of the relatively weak modes. This is noticeable for the weak LP02 (Fig.~\ref{Figure3}(e), red column) and also evident in the green column where both Figs.~\ref{Figure3}(d,e) are likely the weak, leaking LP21 modes. Nevertheless, despite these noise-induced effects, the mixed state reconstruction in Fig.~\ref{Figure3}(f), obtained as a probability weighted intensity sum of all 21 spatial eigenstates, is in almost perfect match with the intensity profiles (Fig.~\ref{Figure3}(g)) obtained by raster scan, which implies accurate determination of the spatial modes and their relative powers. We perform the raster scan by deflecting the beam using linear phase ramps (tilt) at the SLM, which scans the beam around the fixed core of the single-mode fibre. For each tilt, we collect the corresponding spectrum to generate the $(x,y,\lambda)$ datacube, with relevant wavelength slices plotted in Fig.~\ref{Figure3}(g). The influence of the noise level on the Stokes analysis is explored in-depth in Supplementary Note 9 for different levels of noise, RLG basis dimensionality, for pure and mixed states and also exploring the effect of orthogonality of input spatial states.

\subsection*{Spatio-temporal analysis}
\begin{figure*}[ht!]
	\includegraphics [width=16.0cm] {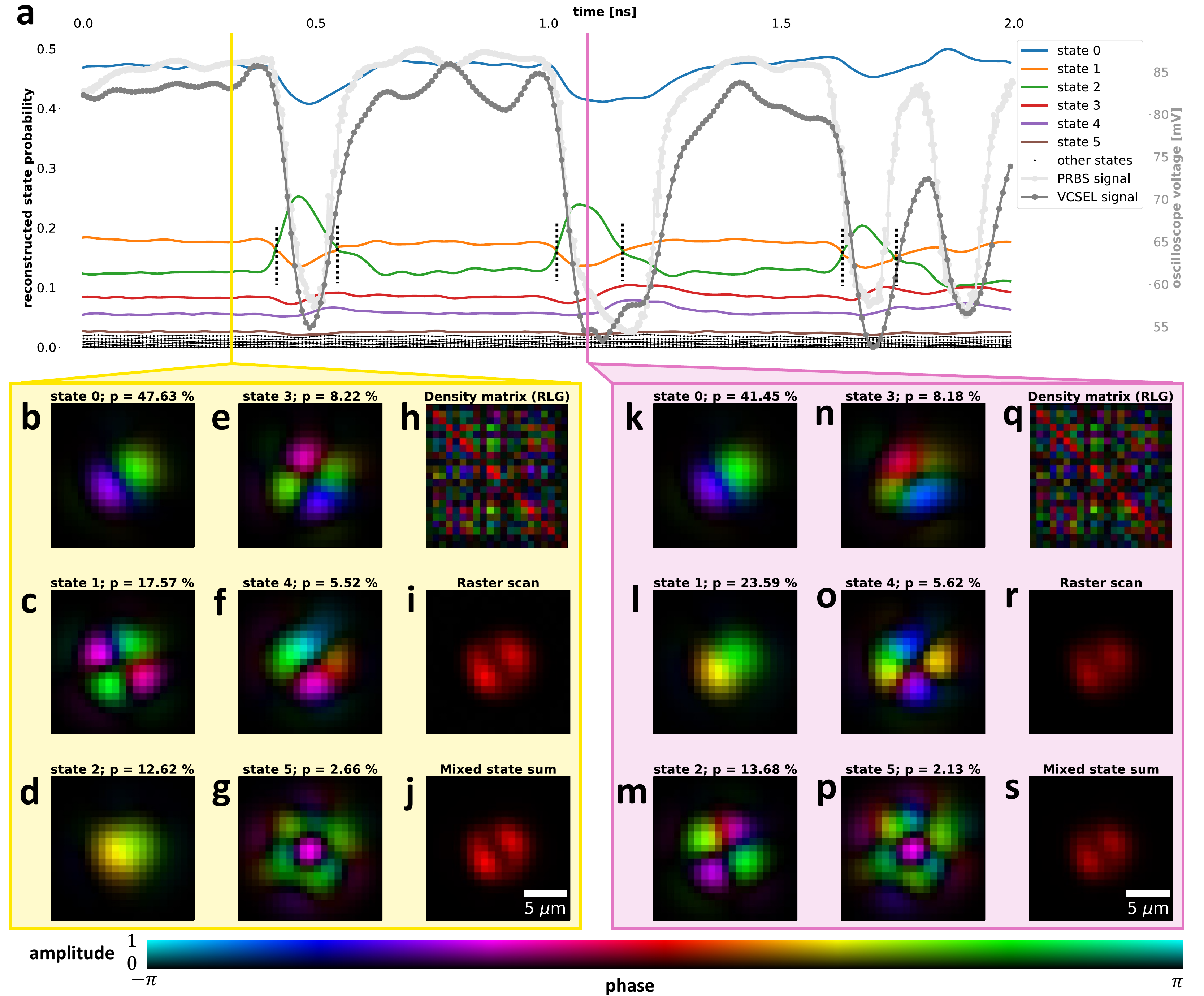}
	\caption{\textbf{Spatio-temporal analysis of VCSEL (H polarisation)} (\textbf{a}) Probability of the reconstructed states (coloured lines) dynamically evolves during modulation (light-grey line) and differs from the overall VCSEL response (dark-grey line). (\textbf{b-g}) and (\textbf{k-p}) are the first six reconstructed spatial eigenstates of density matrices (\textbf{h}) and (\textbf{q}) acquired at times $t = 0.32\,{\rm ns}$ and $t = 1.08\,{\rm ns}$ respectively. (\textbf{i}) and (\textbf{r}) are the total spatial intensity profiles measured by the SLM raster-scan at times $t = 0.32\,{\rm ns}$ and $t = 1.08\,{\rm ns}$ respectively. (\textbf{j}) and (\textbf{s}) are the total spatial intensity profiles calculated as a probability-weighted sum of intensity profiles of all reconstructed spatial states at times $t = 0.32\,{\rm ns}$ and $t = 1.08\,{\rm ns}$ respectively. The black dashed vertical lines mark the probability crossing points between eigenstate 1 and 2.  For the behaviour of the system over the whole $2\,{\rm ns}$ interval see Supplementary Media 3. The V polarisation case is studied in Supplementary Note 6 and Supplementary Media 4.}   
	\label{Figure4}
\end{figure*}
The number of spatial modes, their spectral peak position and relative power depends on the VCSEL temperature and bias. The applied bias does not have to be constant. VCSELs are often modulated at speeds of tens of GHz in short-reach interconnects that form the backbone of modern data centers and local area networks. The quality and speed of modulation can be influenced by mode competition within the cavity and the dynamic modal content of the beam impacts the data transmission through multimode fibres due to modal and chromatic dispersion. Having access to the temporal dynamics of VCSELs on an individual spatial mode basis is therefore of paramount importance in the quest to design faster optical interconnects.

The measured temporally resolved density matrix distinguishes individual mutually incoherent modes, all evolving simultaneously at high speed within the beam. In the spatio-spectral case, the method distinguished between mutually incoherent modes residing in the same spectral bin. Here, the method resolves between all mutually incoherent fields of the beam in each temporal bin, which allows tracking their power (probability) as a function of time.  The ability of the method to differentiate mutually incoherent fields despite no way of separating them in time or spectrum is the added layer of flexibility that empowers the method. Whenever there is a lack of temporal or spectral resolution, the method can still spatially resolve the fields, which can be seen as a form of spontaneous spatial filtering. 

We modulate the VCSEL via a pseudorandom binary sequence (PRBS) generator that is synchronised with an oscilloscope (more details of optical setup in the Supplementary Note 1). The PRBS enables repeatable temporal modulation of the VCSEL between the high ($7\,{\rm mA}$) and the low ($5\,{\rm mA}$) bias states. PRBS modulation in tandem with a synchronised oscilloscope facilitates the sequential application of several hundreds of spatial analyser states on the optical output with identical temporal modulation. We emphasize that the SLM refresh rate does not limit our temporal resolution as we collect the full temporal trace at $22\,{\rm GHz}$ sampling rate -- given by the bandwidth of the photodiode used -- for each spatial analyser state displayed on the SLM. 

The modulation of VCSEL disrupts the equilibrium in the cavity, which triggers dynamic changes to modes, with their relative power evolving and redistributing. Our technique allows observation of these changes for all spatial mode fields, at all delays. Fig.~\ref{Figure4} shows two temporal snapshots of the VCSEL at times $t = 0.32\,{\rm ns}$ (yellow box, high bias) and $t = 1.08\,{\rm ns}$ (purple box, low bias). The selected times correspond to the opposite extremes of bias applied to the VCSEL with values of $7\,{\rm mA}$ and $5\,{\rm mA}$ respectively. The overall VCSEL response (dark-grey curve, obtained from the SLM raster scan and integrated spatially for each temporal bin) closely follows the PRBS drive signal (light-grey curve) modulating the VCSEL at $10\, {\rm GHz}$. The remaining coloured curves in Fig.~\ref{Figure4}(a) are the probabilities of the spatial states that are given by the eigenvectors of the density matrices (Fig.~\ref{Figure4}(h,q)) in a given temporal bin. 

The reconstructed spatial profiles of the first six dominant states are in Figs.~\ref{Figure4}(b-g, k-p). Interestingly, during each transition from high to low bias state and vice-versa, the probability distribution of the spatial eigenstates change. The observed change is most evident on the LP21 (Fig.~\ref{Figure4}(c)) and LP01 (Fig.~\ref{Figure4}(d)) eigenstates. For high bias state, LP21 (Fig.~\ref{Figure4}(c)) dominates over LP01 (Fig.~\ref{Figure4}(d)) by about $5\%$. The situation is completely reversed for the low bias case, with LP01 (Fig.~\ref{Figure4}(l)) having $10\%$ higher probability compared to LP21 (Fig.~\ref{Figure4}(m)). This is an expected behaviour for laser diodes, with low bias favouring the low order modes, in this case the fundamental mode LP01, at the expense of higher order modes. Due to this effect, the overall spatial intensity profile of VCSEL is not constant during modulation as can be seen by comparing Figs.~\ref{Figure4}(i) and Figs.~\ref{Figure4}(r) obtained by SLM raster scan. The high-bias intensity profile in Fig.~\ref{Figure4}(i) is brighter and has more pronounced side lobes due to the higher probability of LP11 and LP21 modes compared to the low bias case in Fig.~\ref{Figure4}(r) which is not only dimmer but also more influenced by the LP01 mode. The observed eigenstate probability changes as a function of bias and follows similar trends to measurements of modally resolved light-current curves presented in Supplementary Note 7. We also note that the SLM raster scan profiles (Figs.~\ref{Figure4}(i,r)) are in agreement with the probability-weighted sum of intensity profiles of all reconstructed spatial states (Figs.~\ref{Figure4}(j,s)), obtained using the high-dimensional Stokes method. This manifests the viability of the method for temporal analysis of the beam.

Compared to the spatio-spectral analysis case, the reconstructed profiles of spatial modes appear to have distorted wavefronts. To determine the cause of this effect, we measured the variation of the VCSEL power for a fixed SLM mask (around $5\%$) and performed the numerical simulations of the Stokes analysis for a mixture of modes with a relative power given by the relative strength of the spectral peaks in Fig.~\ref{Figure3}(a). A total of $6$ modes detected in the spectral analysis (the LP31a and LP31b are not counted as they are weak) is received by the temporal detector at all times, which means that the Stokes analysis has to distinguish a mixture of $6$ modes for each temporal bin (compared to 2 or 3 in spectral analysis). The numerical simulation yields similar wavefront distortion of the reconstructed modes, with increasing level of noise negatively affecting the field overlap value between the input and the reconstructed modes (see Supplementary Note 9 for detailed analysis of the noise effect on Stokes analysis).  An additional source of noise can be attributed to the spatial modes of the real-world cavity being not completely orthogonal due to current injection patterns and manufacturing defects. This is evidenced by LP11 odd and even modes not having perfectly orthogonal orientation of lobes in SLM raster scans of Fig.~\ref{Figure3}(g). In spectral analysis case, the effect of this imperfect orthogonality on wavefront reconstruction is not so evident because of limited number of modes in each spectral bin compared to the temporal case. We numerically simulate the effect of imperfect orthogonality on reconstructed modes in Supplementary Note 9. 

The reconstructed modes and their temporally resolved probabilities/relative powers provide unique insight into the temporal dynamics within the cavity on a mode-by-mode basis. This result can be useful to explore VCSELs of varying aperture scale and shape to develop higher yield of optimum mode dynamics during fabrication. The typically used raster-scanning approaches only obtain the intensity variation of the VCSEL during modulation similar to Figs.~\ref{Figure4}(i,r). While such information is useful, it does not elucidate the role of individual modes. Our tomographic approach provides not only the overall intensity profile (Fig.~\ref{Figure4}(j,s)), but crucially,  the amplitude, phase and the relative powers of all spatial modes at each time.

\section*{Discussion}
The demonstrated technique takes a major step towards enabling characterisation of all the classical degrees of freedom of light -- its spatial, spectral and polarisation components, as well as their temporal evolution. The approach constitutes a simple add-on system that can analyse an arbitrary beam as it is, without access to auxiliary light sources. Moreover, as the method leverages all possible internal references, it is able to analyse even incoherent light sources where there is no suitable self-reference. Interestingly, the technique also distinguishes spatial phase, amplitude and coherence of multiple mutually incoherent arbitrary fields within a single beam, including in situations when these fields spectrally overlap or have the same time delay.  All this information is acquired in a remarkably compressed way -- using a single-pixel detector in the form of single-mode fibre and intensity-only measurements. No judicious choice of spatial basis is required as the outcome of the method is basis-agnostic, which we demonstrated by using a random spatial basis throughout this manuscript. Finally, the method works for any wavelength band for which the spatial light modulator, spectrometer and photodiode exist - with most photonics labs worldwide equipped to implement the technique at almost no additional cost. 

The proof-of-concept experiments explored the intricate spatiotemporal and spectral output of a modulated vertical-cavity surface-emitting laser (VCSEL) diode that represents a class of optical beams difficult to analyse using existing techniques. The lack of complete characterisation of such laser sources represents a bottleneck in a growing number of consumer, medical, automotive, sensing and imaging applications\cite{ebeling_vertical-cavity_2018}. Our method unlocks information that can benefit these applications, as well as support many other areas where the full knowledge of the beam is critical, such as laser engineering \cite{bittner_suppressing_2018}, optical communication in fibres and free-space\cite{zhu_compensation-free_2021}, quantum information\cite{padgett_lights_2014} and remote sensing\cite{belmonte_measurement_2015}. Moreover, the compatibility of the method with any intensity-based detectors opens diverse opportunities for in-depth laser diagnostics. For example, if radio-frequency spectrum analyser is used, the method can recover modally resolved relative intensity noise. The noise reduction is an important aspect of VCSEL engineering that plays a critical role in endeavours to develop $100\,{\rm GHz}$ VCSELs for next-generation interconnects. Finally, based on the knowledge of the modal content, it is possible to design spatial filters that can be either used to launch one specific mode or enable observation of a target spatial mode in isolation from the rest of the system. We demonstrate these applications in Supplementary Note 10. Such spatial filter toolbox can be indispensable for establishing and removing "rogue" modes that impair the overall performance of the system -- for example, from the modulation speed or the relative intensity noise perspective. 

The next-generation design of the method will aim to provide real-time feedback on the optical beam, which is not supported in the current configuration due to the sequential nature of the projection measurements. This problem can be alleviated either by utilising multiplexed holograms \cite{kaiser_complete_2009} or by using multiplane-light-converters\cite{fontaine_laguerre-gaussian_2019, carpenter_optical_2020} that both allow simultaneous interrogation of light with all spatial analyser states in parallel. The detection in such a scenario can be realised via a single-photon avalanche diode array (time) and a hyperspectral camera (spectrum). Additional improvements can be implemented in polarisation detection. Currently, our proof-of-concept system can only infer global polarisation characteristics across the whole cross-section of the beam. Extending the method to recover local polarisation information is possible by using the spatial and polarisation analysers in tandem, with both types of analysers implemented on respective SLMs\cite{moreno_complete_2012} to create the required projection states. Such modification would make our method suitable for the analysis of vector beams \cite{rosales-guzman_review_2018}, where the polarisation depends on the spatial position.

\vspace{0.2cm}
\textbf{Funding and Acknowledgements}
\vspace{0.2cm}

{\footnotesize This work has been supported by the Australian Research Council (ARC) funding through ARC Linkage (LP170100720). M.P., M.M. and J.C. acknowledge support through Discovery Early Career Research Awards (DE170100241, DE210100934 and DE180100009). We acknowledge Yah Leng Lim from The University of Queensland and Frank Flens from II-VI Incorporated for helpful discussions.}

\vspace{0.2cm}
\textbf{Author contributions}
\vspace{0.2cm}

{\footnotesize M.P. designed and built the instrumentation, performed the experiments, analysed the data and developed the numerical framework. M.M.M,  D.D., M.M. and J.C. assisted M.P. with the analysis of the spatio-temporal and spatio-spectral results and contributed to the numerical implementation of the high-dimensional Stokes tomography developed by M.P. G.L. assisted M.P. with driving the VCSEL system. G.L. also assisted M.P with the interpretation of the light-current experiments, with auxiliary input from A.R. M.P. wrote the manuscript with contributions from J.C., M.M., M.M.M and D.D. J.C. conceived the idea and co-led the project with M.P.}

\vspace{0.2cm}
\textbf{Disclosures}
\vspace{0.2cm}

{\footnotesize G.L. is employed by II-VI Incorporated. The authors declare that they have no known competing financial interests or personal relationships that could have appeared to influence the work reported in this paper.}

\vspace{0.2cm}

\textbf{Data availability}
\vspace{0.2cm}

{\footnotesize The data that support the findings of this study and the software to obtain and analyse the results are available from the corresponding author upon reasonable request.}

\newpage
\bibliographystyle{Naturemag}

\bibliography{VCSEL_project}

\begin{thebibliography}{10}
\expandafter\ifx\csname url\endcsname\relax
  \def\url#1{\texttt{#1}}\fi
\expandafter\ifx\csname urlprefix\endcsname\relax\def\urlprefix{URL }\fi
\providecommand{\bibinfo}[2]{#2}
\providecommand{\eprint}[2][]{\url{#2}}

\bibitem{mosk_controlling_2012}
\bibinfo{author}{Mosk, A.~P.}, \bibinfo{author}{Lagendijk, A.},
  \bibinfo{author}{Lerosey, G.} \& \bibinfo{author}{Fink, M.}
\newblock \bibinfo{title}{Controlling waves in space and time for imaging and
  focusing in complex media}.
\newblock \emph{\bibinfo{journal}{Nature Photon}} \textbf{\bibinfo{volume}{6}},
  \bibinfo{pages}{283--292} (\bibinfo{year}{2012}).

\bibitem{dholakia_shaping_2011}
\bibinfo{author}{Dholakia, K.} \& \bibinfo{author}{Čižmár, T.}
\newblock \bibinfo{title}{Shaping the future of manipulation}.
\newblock \emph{\bibinfo{journal}{Nature Photon}} \textbf{\bibinfo{volume}{5}},
  \bibinfo{pages}{335--342} (\bibinfo{year}{2011}).

\bibitem{rubinsztein-dunlop_roadmap_2017}
\bibinfo{author}{Rubinsztein-Dunlop, H.} \emph{et~al.}
\newblock \bibinfo{title}{Roadmap on structured light}.
\newblock \emph{\bibinfo{journal}{J. Opt.}} \textbf{\bibinfo{volume}{19}},
  \bibinfo{pages}{013001} (\bibinfo{year}{2017}).

\bibitem{weiner_femtosecond_2000}
\bibinfo{author}{Weiner, A.~M.}
\newblock \bibinfo{title}{Femtosecond pulse shaping using spatial light
  modulators}.
\newblock \emph{\bibinfo{journal}{Review of Scientific Instruments}}
  \textbf{\bibinfo{volume}{71}}, \bibinfo{pages}{1929--1960}
  (\bibinfo{year}{2000}).

\bibitem{brejnak_boosting_2021}
\bibinfo{author}{Brejnak, A.} \emph{et~al.}
\newblock \bibinfo{title}{Boosting the output power of large-aperture lasers by
  breaking their circular symmetry}.
\newblock \emph{\bibinfo{journal}{Optica}} \textbf{\bibinfo{volume}{8}},
  \bibinfo{pages}{1167} (\bibinfo{year}{2021}).

\bibitem{gensty_wave_2005}
\bibinfo{author}{Gensty, T.} \emph{et~al.}
\newblock \bibinfo{title}{Wave {Chaos} in {Real}-{World} {Vertical}-{Cavity}
  {Surface}-{Emitting} {Lasers}}.
\newblock \emph{\bibinfo{journal}{Phys. Rev. Lett.}}
  \textbf{\bibinfo{volume}{94}}, \bibinfo{pages}{233901}
  (\bibinfo{year}{2005}).

\bibitem{bittner_suppressing_2018}
\bibinfo{author}{Bittner, S.} \emph{et~al.}
\newblock \bibinfo{title}{Suppressing spatiotemporal lasing instabilities with
  wave-chaotic microcavities}.
\newblock \emph{\bibinfo{journal}{Science}} \textbf{\bibinfo{volume}{361}},
  \bibinfo{pages}{1225--1231} (\bibinfo{year}{2018}).

\bibitem{booth_adaptive_2014}
\bibinfo{author}{Booth, M.~J.}
\newblock \bibinfo{title}{Adaptive optical microscopy: the ongoing quest for a
  perfect image}.
\newblock \emph{\bibinfo{journal}{Light Sci Appl}}
  \textbf{\bibinfo{volume}{3}}, \bibinfo{pages}{e165--e165}
  (\bibinfo{year}{2014}).

\bibitem{mounaix_time_2020}
\bibinfo{author}{Mounaix, M.} \emph{et~al.}
\newblock \bibinfo{title}{Time reversed optical waves by arbitrary vector
  spatiotemporal field generation}.
\newblock \emph{\bibinfo{journal}{Nat Commun}} \textbf{\bibinfo{volume}{11}},
  \bibinfo{pages}{5813} (\bibinfo{year}{2020}).

\bibitem{wright_controllable_2015}
\bibinfo{author}{Wright, L.~G.}, \bibinfo{author}{Christodoulides, D.~N.} \&
  \bibinfo{author}{Wise, F.~W.}
\newblock \bibinfo{title}{Controllable spatiotemporal nonlinear effects in
  multimode fibres}.
\newblock \emph{\bibinfo{journal}{Nature Photon}} \textbf{\bibinfo{volume}{9}},
  \bibinfo{pages}{306--310} (\bibinfo{year}{2015}).

\bibitem{wright_spatiotemporal_2017}
\bibinfo{author}{Wright, L.~G.}, \bibinfo{author}{Christodoulides, D.~N.} \&
  \bibinfo{author}{Wise, F.~W.}
\newblock \bibinfo{title}{Spatiotemporal mode-locking in multimode fiber
  lasers}.
\newblock \emph{\bibinfo{journal}{Science}} \textbf{\bibinfo{volume}{358}},
  \bibinfo{pages}{94--97} (\bibinfo{year}{2017}).

\bibitem{nicholson_spatially_2008}
\bibinfo{author}{Nicholson, J.~W.}, \bibinfo{author}{Yablon, A.~D.},
  \bibinfo{author}{Ramachandran, S.} \& \bibinfo{author}{Ghalmi, S.}
\newblock \bibinfo{title}{Spatially and spectrally resolved imaging of modal
  content in large-mode-area fibers}.
\newblock \emph{\bibinfo{journal}{Opt. Express}} \textbf{\bibinfo{volume}{16}},
  \bibinfo{pages}{7233} (\bibinfo{year}{2008}).

\bibitem{gao_single-shot_2014}
\bibinfo{author}{Gao, L.}, \bibinfo{author}{Liang, J.}, \bibinfo{author}{Li,
  C.} \& \bibinfo{author}{Wang, L.~V.}
\newblock \bibinfo{title}{Single-shot compressed ultrafast photography at one
  hundred billion frames per second}.
\newblock \emph{\bibinfo{journal}{Nature}} \textbf{\bibinfo{volume}{516}},
  \bibinfo{pages}{74--77} (\bibinfo{year}{2014}).

\bibitem{yamaguchi_phase-shifting_1997}
\bibinfo{author}{Yamaguchi, I.} \& \bibinfo{author}{Zhang, T.}
\newblock \bibinfo{title}{Phase-shifting digital holography}.
\newblock \emph{\bibinfo{journal}{Opt. Lett.}} \textbf{\bibinfo{volume}{22}},
  \bibinfo{pages}{1268} (\bibinfo{year}{1997}).

\bibitem{goodman_digital_1967}
\bibinfo{author}{Goodman, J.~W.} \& \bibinfo{author}{Lawrence, R.~W.}
\newblock \bibinfo{title}{Digital image formation from electronically detected
  holograms}.
\newblock \emph{\bibinfo{journal}{Appl. Phys. Lett.}}
  \textbf{\bibinfo{volume}{11}}, \bibinfo{pages}{77--79}
  (\bibinfo{year}{1967}).

\bibitem{fontaine_laguerre-gaussian_2019}
\bibinfo{author}{Fontaine, N.~K.} \emph{et~al.}
\newblock \bibinfo{title}{Laguerre-{Gaussian} mode sorter}.
\newblock \emph{\bibinfo{journal}{Nat Commun}} \textbf{\bibinfo{volume}{10}},
  \bibinfo{pages}{1865} (\bibinfo{year}{2019}).

\bibitem{kaiser_complete_2009}
\bibinfo{author}{Kaiser, T.}, \bibinfo{author}{Flamm, D.},
  \bibinfo{author}{Schröter, S.} \& \bibinfo{author}{Duparré, M.}
\newblock \bibinfo{title}{Complete modal decomposition for optical fibers using
  {CGH}-based correlation filters}.
\newblock \emph{\bibinfo{journal}{Opt. Express}} \textbf{\bibinfo{volume}{17}},
  \bibinfo{pages}{9347} (\bibinfo{year}{2009}).

\bibitem{forbes_creation_2016}
\bibinfo{author}{Forbes, A.}, \bibinfo{author}{Dudley, A.} \&
  \bibinfo{author}{McLaren, M.}
\newblock \bibinfo{title}{Creation and detection of optical modes with spatial
  light modulators}.
\newblock \emph{\bibinfo{journal}{Adv. Opt. Photon.}}
  \textbf{\bibinfo{volume}{8}}, \bibinfo{pages}{200} (\bibinfo{year}{2016}).

\bibitem{pinnell_modal_2020}
\bibinfo{author}{Pinnell, J.} \emph{et~al.}
\newblock \bibinfo{title}{Modal analysis of structured light with spatial light
  modulators: a practical tutorial}.
\newblock \emph{\bibinfo{journal}{J. Opt. Soc. Am. A}}
  \textbf{\bibinfo{volume}{37}}, \bibinfo{pages}{C146} (\bibinfo{year}{2020}).

\bibitem{flamm_mode_2012}
\bibinfo{author}{Flamm, D.}, \bibinfo{author}{Naidoo, D.},
  \bibinfo{author}{Schulze, C.}, \bibinfo{author}{Forbes, A.} \&
  \bibinfo{author}{Duparré, M.}
\newblock \bibinfo{title}{Mode analysis with a spatial light modulator as a
  correlation filter}.
\newblock \emph{\bibinfo{journal}{Opt. Lett.}} \textbf{\bibinfo{volume}{37}},
  \bibinfo{pages}{2478} (\bibinfo{year}{2012}).

\bibitem{schulze_wavefront_2012}
\bibinfo{author}{Schulze, C.} \emph{et~al.}
\newblock \bibinfo{title}{Wavefront reconstruction by modal decomposition}.
\newblock \emph{\bibinfo{journal}{Opt. Express}} \textbf{\bibinfo{volume}{20}},
  \bibinfo{pages}{19714} (\bibinfo{year}{2012}).

\bibitem{paurisse_complete_2012}
\bibinfo{author}{Paurisse, M.}, \bibinfo{author}{Lévèque, L.},
  \bibinfo{author}{Hanna, M.}, \bibinfo{author}{Druon, F.} \&
  \bibinfo{author}{Georges, P.}
\newblock \bibinfo{title}{Complete measurement of fiber modal content by
  wavefront analysis}.
\newblock \emph{\bibinfo{journal}{Opt. Express}} \textbf{\bibinfo{volume}{20}},
  \bibinfo{pages}{4074} (\bibinfo{year}{2012}).

\bibitem{stutzki_high-speed_2011}
\bibinfo{author}{Stutzki, F.} \emph{et~al.}
\newblock \bibinfo{title}{High-speed modal decomposition of mode instabilities
  in high-power fiber lasers}.
\newblock \emph{\bibinfo{journal}{Opt. Lett.}} \textbf{\bibinfo{volume}{36}},
  \bibinfo{pages}{4572} (\bibinfo{year}{2011}).

\bibitem{duparre_-line_2005}
\bibinfo{author}{Duparré, M.}, \bibinfo{author}{Lüdge, B.} \&
  \bibinfo{author}{Schröter, S.}
\newblock \bibinfo{title}{On-line characterization of {Nd}:{YAG} laser beams by
  means of modal decomposition using diffractive optical correlation filters}.
\newblock \bibinfo{pages}{59622G} (\bibinfo{address}{Jena, Germany},
  \bibinfo{year}{2005}).

\bibitem{schmidt_real-time_2011}
\bibinfo{author}{Schmidt, O.~A.} \emph{et~al.}
\newblock \bibinfo{title}{Real-time determination of laser beam quality by
  modal decomposition}.
\newblock \emph{\bibinfo{journal}{Opt. Express}} \textbf{\bibinfo{volume}{19}},
  \bibinfo{pages}{6741} (\bibinfo{year}{2011}).

\bibitem{lvovsky_continuous-variable_2009}
\bibinfo{author}{Lvovsky, A.~I.} \& \bibinfo{author}{Raymer, M.~G.}
\newblock \bibinfo{title}{Continuous-variable optical quantum-state
  tomography}.
\newblock \emph{\bibinfo{journal}{Rev. Mod. Phys.}}
  \textbf{\bibinfo{volume}{81}}, \bibinfo{pages}{299--332}
  (\bibinfo{year}{2009}).

\bibitem{toninelli_concepts_2019}
\bibinfo{author}{Toninelli, E.} \emph{et~al.}
\newblock \bibinfo{title}{Concepts in quantum state tomography and classical
  implementation with intense light: a tutorial}.
\newblock \emph{\bibinfo{journal}{Adv. Opt. Photon.}}
  \textbf{\bibinfo{volume}{11}}, \bibinfo{pages}{67} (\bibinfo{year}{2019}).

\bibitem{dunn_experimental_1995}
\bibinfo{author}{Dunn, T.~J.}, \bibinfo{author}{Walmsley, I.~A.} \&
  \bibinfo{author}{Mukamel, S.}
\newblock \bibinfo{title}{Experimental {Determination} of the
  {Quantum}-{Mechanical} {State} of a {Molecular} {Vibrational} {Mode} {Using}
  {Fluorescence} {Tomography}}.
\newblock \emph{\bibinfo{journal}{Phys. Rev. Lett.}}
  \textbf{\bibinfo{volume}{74}}, \bibinfo{pages}{884--887}
  (\bibinfo{year}{1995}).

\bibitem{obrien_demonstration_2003}
\bibinfo{author}{O'Brien, J.~L.}, \bibinfo{author}{Pryde, G.~J.},
  \bibinfo{author}{White, A.~G.}, \bibinfo{author}{Ralph, T.~C.} \&
  \bibinfo{author}{Branning, D.}
\newblock \bibinfo{title}{Demonstration of an all-optical quantum
  controlled-{NOT} gate}.
\newblock \emph{\bibinfo{journal}{Nature}} \textbf{\bibinfo{volume}{426}},
  \bibinfo{pages}{264--267} (\bibinfo{year}{2003}).

\bibitem{mclaren_entangled_2012}
\bibinfo{author}{McLaren, M.} \emph{et~al.}
\newblock \bibinfo{title}{Entangled {Bessel}-{Gaussian} beams}.
\newblock \emph{\bibinfo{journal}{Opt. Express}} \textbf{\bibinfo{volume}{20}},
  \bibinfo{pages}{23589} (\bibinfo{year}{2012}).

\bibitem{yang_using_2016}
\bibinfo{author}{Yang, J.} \& \bibinfo{author}{Nolan, D.~A.}
\newblock \bibinfo{title}{Using state tomography for characterizing input
  principal modes in optically scattering medium}.
\newblock \emph{\bibinfo{journal}{Opt. Express}} \textbf{\bibinfo{volume}{24}},
  \bibinfo{pages}{27691} (\bibinfo{year}{2016}).

\bibitem{milione_determining_2015}
\bibinfo{author}{Milione, G.}, \bibinfo{author}{Nolan, D.~A.} \&
  \bibinfo{author}{Alfano, R.~R.}
\newblock \bibinfo{title}{Determining principal modes in a multimode optical
  fiber using the mode dependent signal delay method}.
\newblock \emph{\bibinfo{journal}{J. Opt. Soc. Am. B}}
  \textbf{\bibinfo{volume}{32}}, \bibinfo{pages}{143} (\bibinfo{year}{2015}).

\bibitem{milione_higher-order_2011}
\bibinfo{author}{Milione, G.}, \bibinfo{author}{Sztul, H.~I.},
  \bibinfo{author}{Nolan, D.~A.} \& \bibinfo{author}{Alfano, R.~R.}
\newblock \bibinfo{title}{Higher-{Order} {Poincaré} {Sphere}, {Stokes}
  {Parameters}, and the {Angular} {Momentum} of {Light}}.
\newblock \emph{\bibinfo{journal}{Phys. Rev. Lett.}}
  \textbf{\bibinfo{volume}{107}}, \bibinfo{pages}{053601}
  (\bibinfo{year}{2011}).

\bibitem{ji_high-dimensional_2019}
\bibinfo{author}{Ji, H.} \emph{et~al.}
\newblock \bibinfo{title}{High-dimensional {Stokes} vector direct detection
  over few-mode fibers}.
\newblock \emph{\bibinfo{journal}{Opt. Lett.}} \textbf{\bibinfo{volume}{44}},
  \bibinfo{pages}{2065} (\bibinfo{year}{2019}).

\bibitem{dennis_swings_2017}
\bibinfo{author}{Dennis, M.~R.} \& \bibinfo{author}{Alonso, M.~A.}
\newblock \bibinfo{title}{Swings and roundabouts: optical {Poincaré} spheres
  for polarization and {Gaussian} beams}.
\newblock \emph{\bibinfo{journal}{Phil. Trans. R. Soc. A.}}
  \textbf{\bibinfo{volume}{375}}, \bibinfo{pages}{20150441}
  (\bibinfo{year}{2017}).

\bibitem{agnew_tomography_2011}
\bibinfo{author}{Agnew, M.}, \bibinfo{author}{Leach, J.},
  \bibinfo{author}{McLaren, M.}, \bibinfo{author}{Roux, F.~S.} \&
  \bibinfo{author}{Boyd, R.~W.}
\newblock \bibinfo{title}{Tomography of the quantum state of photons entangled
  in high dimensions}.
\newblock \emph{\bibinfo{journal}{Phys. Rev. A}} \textbf{\bibinfo{volume}{84}},
  \bibinfo{pages}{062101} (\bibinfo{year}{2011}).

\bibitem{r_sheppard_three-dimensional_2016}
\bibinfo{author}{R.~Sheppard, C.~J.}, \bibinfo{author}{Castello, M.} \&
  \bibinfo{author}{Diaspro, A.}
\newblock \bibinfo{title}{Three-dimensional polarization algebra}.
\newblock \emph{\bibinfo{journal}{J. Opt. Soc. Am. A}}
  \textbf{\bibinfo{volume}{33}}, \bibinfo{pages}{1938} (\bibinfo{year}{2016}).

\bibitem{salvail_full_2013}
\bibinfo{author}{Salvail, J.~Z.} \emph{et~al.}
\newblock \bibinfo{title}{Full characterization of polarization states of light
  via direct measurement}.
\newblock \emph{\bibinfo{journal}{Nature Photon}} \textbf{\bibinfo{volume}{7}},
  \bibinfo{pages}{316--321} (\bibinfo{year}{2013}).

\bibitem{gil-lopez_universal_2021}
\bibinfo{author}{Gil-Lopez, J.} \emph{et~al.}
\newblock \bibinfo{title}{Universal compressive tomography in the
  time-frequency domain}.
\newblock \emph{\bibinfo{journal}{Optica}} \textbf{\bibinfo{volume}{8}},
  \bibinfo{pages}{1296} (\bibinfo{year}{2021}).

\bibitem{edgar_principles_2019}
\bibinfo{author}{Edgar, M.~P.}, \bibinfo{author}{Gibson, G.~M.} \&
  \bibinfo{author}{Padgett, M.~J.}
\newblock \bibinfo{title}{Principles and prospects for single-pixel imaging}.
\newblock \emph{\bibinfo{journal}{Nature Photon}}
  \textbf{\bibinfo{volume}{13}}, \bibinfo{pages}{13--20}
  (\bibinfo{year}{2019}).

\bibitem{gell-mann_symmetries_1962}
\bibinfo{author}{Gell-Mann, M.}
\newblock \bibinfo{title}{Symmetries of {Baryons} and {Mesons}}.
\newblock \emph{\bibinfo{journal}{Phys. Rev.}} \textbf{\bibinfo{volume}{125}},
  \bibinfo{pages}{1067--1084} (\bibinfo{year}{1962}).

\bibitem{liu_vertical-cavity_2019}
\bibinfo{author}{Liu, A.}, \bibinfo{author}{Wolf, P.}, \bibinfo{author}{Lott,
  J.~A.} \& \bibinfo{author}{Bimberg, D.}
\newblock \bibinfo{title}{Vertical-cavity surface-emitting lasers for data
  communication and sensing}.
\newblock \emph{\bibinfo{journal}{Photon. Res.}} \textbf{\bibinfo{volume}{7}},
  \bibinfo{pages}{121} (\bibinfo{year}{2019}).

\bibitem{higham_computing_1988}
\bibinfo{author}{Higham, N.~J.}
\newblock \bibinfo{title}{Computing a nearest symmetric positive semidefinite
  matrix}.
\newblock \emph{\bibinfo{journal}{Linear Algebra and its Applications}}
  \textbf{\bibinfo{volume}{103}}, \bibinfo{pages}{103--118}
  (\bibinfo{year}{1988}).

\bibitem{ebeling_vertical-cavity_2018}
\bibinfo{author}{Ebeling, K.~J.}, \bibinfo{author}{Michalzik, R.} \&
  \bibinfo{author}{Moench, H.}
\newblock \bibinfo{title}{Vertical-cavity surface-emitting laser technology
  applications with focus on sensors and three-dimensional imaging}.
\newblock \emph{\bibinfo{journal}{Jpn. J. Appl. Phys.}}
  \textbf{\bibinfo{volume}{57}}, \bibinfo{pages}{08PA02}
  (\bibinfo{year}{2018}).

\bibitem{zhu_compensation-free_2021}
\bibinfo{author}{Zhu, Z.} \emph{et~al.}
\newblock \bibinfo{title}{Compensation-free high-dimensional free-space optical
  communication using turbulence-resilient vector beams}.
\newblock \emph{\bibinfo{journal}{Nat Commun}} \textbf{\bibinfo{volume}{12}},
  \bibinfo{pages}{1666} (\bibinfo{year}{2021}).

\bibitem{padgett_lights_2014}
\bibinfo{author}{Padgett, M.}
\newblock \bibinfo{title}{Light's twist}.
\newblock \emph{\bibinfo{journal}{Proc. R. Soc. A.}}
  \textbf{\bibinfo{volume}{470}}, \bibinfo{pages}{20140633}
  (\bibinfo{year}{2014}).

\bibitem{belmonte_measurement_2015}
\bibinfo{author}{Belmonte, A.}, \bibinfo{author}{Rosales-Guzmán, C.} \&
  \bibinfo{author}{Torres, J.~P.}
\newblock \bibinfo{title}{Measurement of flow vorticity with helical beams of
  light}.
\newblock \emph{\bibinfo{journal}{Optica}} \textbf{\bibinfo{volume}{2}},
  \bibinfo{pages}{1002} (\bibinfo{year}{2015}).

\bibitem{carpenter_optical_2020}
\bibinfo{author}{Carpenter, J.} \& \bibinfo{author}{Fontaine, N.~K.}
\newblock \bibinfo{title}{Optical single-shot spatial state tomography}.
\newblock In \emph{\bibinfo{booktitle}{14th {Pacific} {Rim} {Conference} on
  {Lasers} and {Electro}-{Optics} ({CLEO} {PR} 2020)}},
  \bibinfo{pages}{C10C\_4} (\bibinfo{publisher}{OSA},
  \bibinfo{address}{Sydney}, \bibinfo{year}{2020}).

\bibitem{moreno_complete_2012}
\bibinfo{author}{Moreno, I.}, \bibinfo{author}{Davis, J.~A.},
  \bibinfo{author}{Hernandez, T.~M.}, \bibinfo{author}{Cottrell, D.~M.} \&
  \bibinfo{author}{Sand, D.}
\newblock \bibinfo{title}{Complete polarization control of light from a liquid
  crystal spatial light modulator}.
\newblock \emph{\bibinfo{journal}{Opt. Express}} \textbf{\bibinfo{volume}{20}},
  \bibinfo{pages}{364} (\bibinfo{year}{2012}).

\bibitem{rosales-guzman_review_2018}
\bibinfo{author}{Rosales-Guzmán, C.}, \bibinfo{author}{Ndagano, B.} \&
  \bibinfo{author}{Forbes, A.}
\newblock \bibinfo{title}{A review of complex vector light fields and their
  applications}.
\newblock \emph{\bibinfo{journal}{J. Opt.}} \textbf{\bibinfo{volume}{20}},
  \bibinfo{pages}{123001} (\bibinfo{year}{2018}).

\end{thebibliography}

\end{document}